\documentclass[pre,twocolumn,aps,superscriptaddress,floatfix]{revtex4}
\usepackage[pdftex]{graphicx,color}
\usepackage{amssymb}
\usepackage{epsfig}
\usepackage{subfigure}
\usepackage{textcomp}
\usepackage{float}
\usepackage{comment}
\usepackage{url}
\usepackage{bm}
\expandafter\let\csname equation*\endcsname\relax
\expandafter\let\csname endequation*\endcsname\relax
\usepackage{amsmath}
\graphicspath{ {./figs/} }
\def \equi#1{\mathrel{\mathop{\kern 0pt\sim}\limits_{#1}}}

\begin{document}
\title{Does Greed Help a Forager Survive?}

\author{U. Bhat} \affiliation{Department of Physics, Boston University, Boston,
  Massachusetts 02215, USA}\affiliation{Santa Fe Institute, 1399 Hyde Park Road, Santa
  Fe, New Mexico 87501, USA}
  
\author{S. Redner} \affiliation{Santa Fe Institute, 1399 Hyde Park Road, Santa
  Fe, New Mexico 87501, USA}

\author{O. B\'enichou} \affiliation{Laboratoire de Physique Th\'eorique de la
  Mati\`ere Condens\'ee (UMR CNRS 7600), Universit\'e Pierre et Marie Curie,
  4 Place Jussieu, 75252 Paris Cedex France}

\begin{abstract}
  We investigate the role of greed on the lifetime of a random-walking
  forager on an initially resource-rich lattice.  Whenever the forager lands
  on a food-containing site, all the food there is eaten and the forager can
  hop $\mathcal{S}$ more steps without food before starving.  Upon reaching
  an empty site, the forager comes one time unit closer to starvation.  The
  forager is also \emph{greedy}---given a choice to move to an empty or to a
  food-containing site in its local neighborhood, the forager moves
  preferentially towards food.  Surprisingly, the forager lifetime varies
  \emph{non-monotonically} with greed, with different senses of the
  non-monotonicity in one and two dimensions.  Also unexpectedly, the forager
  lifetime in one dimension has a huge peak for very negative greed where the
  forager is food averse.

\end{abstract}
\maketitle

\section{Introduction}

Optimal foraging theory is a classic framework that specifies when a forager
should continue to exploit local resources or move to new feeding
grounds~\cite{C76,KR85,SK86,OB90,B91,ASD97,KM01}.  The goal is to formulate a
strategy to consume the maximal amount of resource per unit time.  Optimal
strategies typically involve the interplay between continuing to exploit
resources in a current search domain or moving to another and potentially
richer search domain.  This same tension underlies a diverse range of
decision-making problems, including, for example, the management of
firms~\cite{M91,GDB14}, the multiarm bandit problem~\cite{R52,G79}, the
secretary problem~\cite{F89} and its variant, Feynman's restaurant
problem~\cite{Fe}, and search of human memory~\cite{HJT12,AAG15}.  These
problems offer a rich arena for applying statistical physics ideas.  An
independent approach to foraging is to search using exotic search strategies,
such as L\'evy walks~\cite{Viswanathan:1999a}, intermittent
walks~\cite{Benichou:2005,Oshanin:2007,Lomholt:2008,Bressloff:2011,VLRS11,PCM14}
and persistent random walks~\cite{Tejedor:2012}.  However, these models
typically do not account for resource depletion in an explicit way.

In the context of resource foraging, we recently introduced the
\emph{starving random walk} model, in which the forager is unaffected by the
presence or absence of food and always performs an unbiased random
walk~\cite{BR14,CBR16}.  When a forager lands on a food-containing site, all
the food there is consumed.  Immediately afterwards, the forager is in a
fully sated state and can hop $\mathcal{S}$ additional steps without again
encountering food before it starves.  However, if the forager lands on an
empty site, the forager goes hungry and comes one time unit closer to
starvation.  Because there is no replenishment, resources are depleted by
consumption and the forager is doomed to ultimately starve to death.  This
feature of depletion makes the forager motion a non-trivial non-Markovian
process.  How does the forager lifetime $\mathcal{T}$ depend on basic
parameters---its metabolic capacity $\mathcal{S}$ and the spatial dimension
$d$?  While there has been progress in answering this
question~\cite{BR14,CBR16}, a full understanding is still incomplete.

\begin{figure}[ht]
\centerline{
\includegraphics[width=0.18\textwidth]{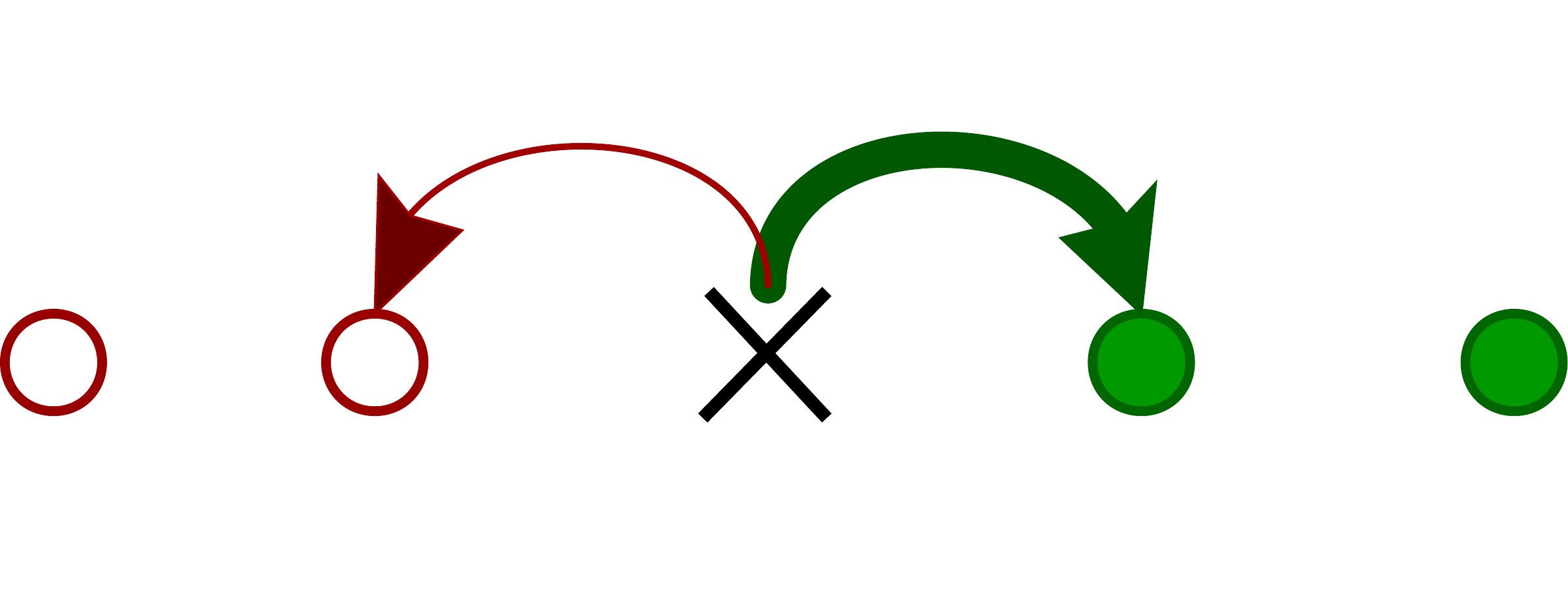}\qquad\qquad
\includegraphics[width=0.18\textwidth]{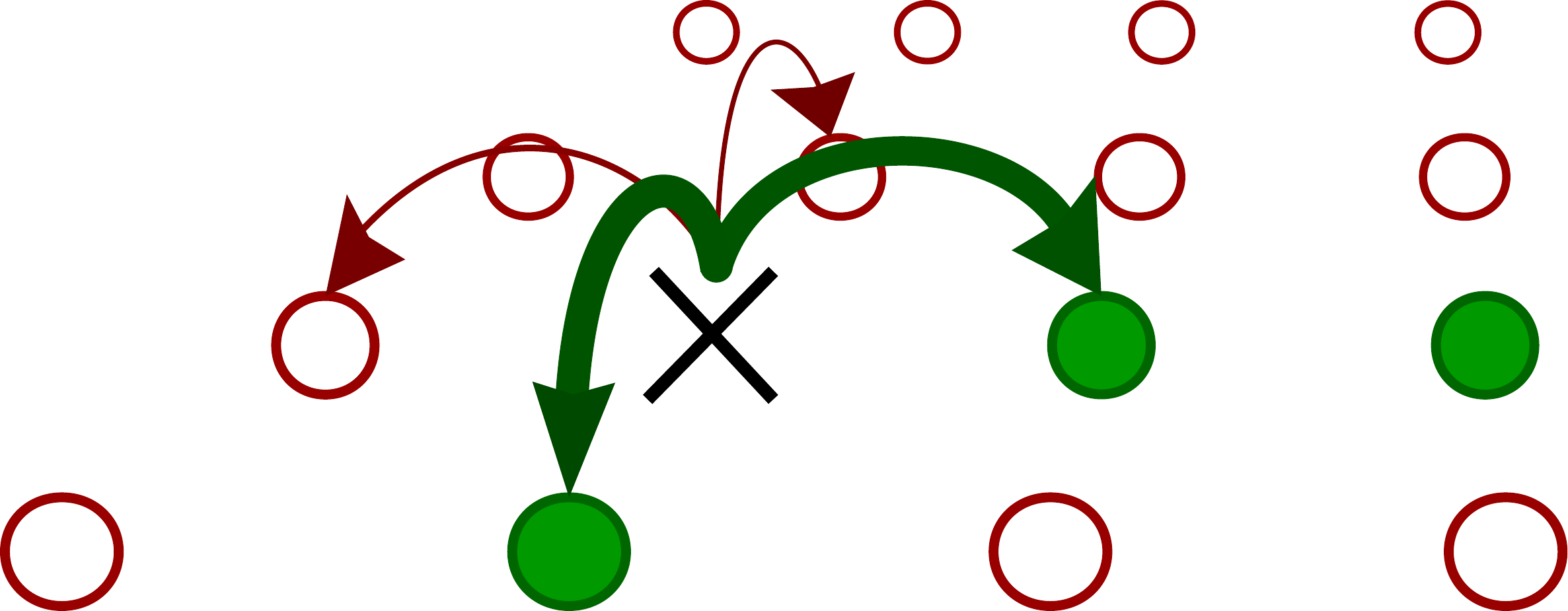}} 
    \caption{ Greedy forager motion in $d=1$ and $d=2$.  Solid and open circles indicate food
      and empty sites.  Arrow widths indicate relative hopping
      probabilities.}
\label{cartoon}
\end{figure}

In this work, we investigate an ecologically motivated extension of the
starving random walk where the forager possesses a modicum of environmental
awareness---whenever the nearest neighborhood of a forager contains both
empty and full (food-containing) sites, the forager preferentially moves
towards the food (Fig.~\ref{cartoon}).  We define this local propensity to
move towards food as ``greed''.  We will also investigate \emph{negative}
greed, or equivalently, food aversion, in which a forager tends to avoid food
in its nearest neighborhood.

Because greed is a universal attribute, its role in optimization processes
has been widely investigated.  In computer science, greedy algorithms are
often an initial approach to solve complex problems~\cite{CLRN01,BGY04,MM11}.
Such algorithms work well for finding the minimal spanning tree of a
graph~\cite{K56} or the ground state of a spin glass~\cite{NS94}, but work
less well for the traveling salesman problem~\cite{MM11} and depth first
search processes~\cite{G}.  Greed also represents a particularly simple
example of feedback between the environmental state and the forager motion, a
mechanism that abounds in the microscopic world.  Perhaps the best-known
example is the run and tumble model of chemotaxis~\cite{BB72,BP77,DZ88}, in
which a bacterium effectively swims up a concentration gradient of
nourishment.  In chemotaxis, however, the concentration of nutrients is
fixed, while the starving forager model explicitly incorporates resource
depletion.

Endowing a starving random walker with greed allows us to discuss the
dichotomy between exploration and exploitation in foraging problems---should
one continue to exploit a rich local lode in a ``desert'' or is it better to
move to a region where resources are more abundant
overall~\cite{BR08,OCF13,PBHIW}?  This is the basic question that we address
by extending first-passage techniques to the unconventional random walk that
arises because of the local bias whenever the forager encounters food.

In $d=1$, we implement greed as follows: when one neighbor of the forager
contains food while the other is empty, the forager moves towards the food
with probability $p=(1+G)/2$, where $G$ is the greediness parameter that lies
in $[-1,1]$; otherwise, the forager hops symmetrically (Fig.~\ref{cartoon}).
For $d>1$ the forager chooses one of the $k$ full sites in its neighborhood
of $z$ sites with probability $p={(1+G)}/{\big[(z-k)(1-G)+k(1+G)\big]}$.  The
forager begins in the ``Eden'' condition where all sites initially contain
food.  As the forager moves, it carves out a food-depleted region---the
``desert''.  As this desert grows, the forager typically spends longer times
wandering within the desert and eventually starves.

\section{Heuristics For One Dimension} 

\begin{figure}[ht]
\hspace{0.35cm}\includegraphics[width=0.4\textwidth]{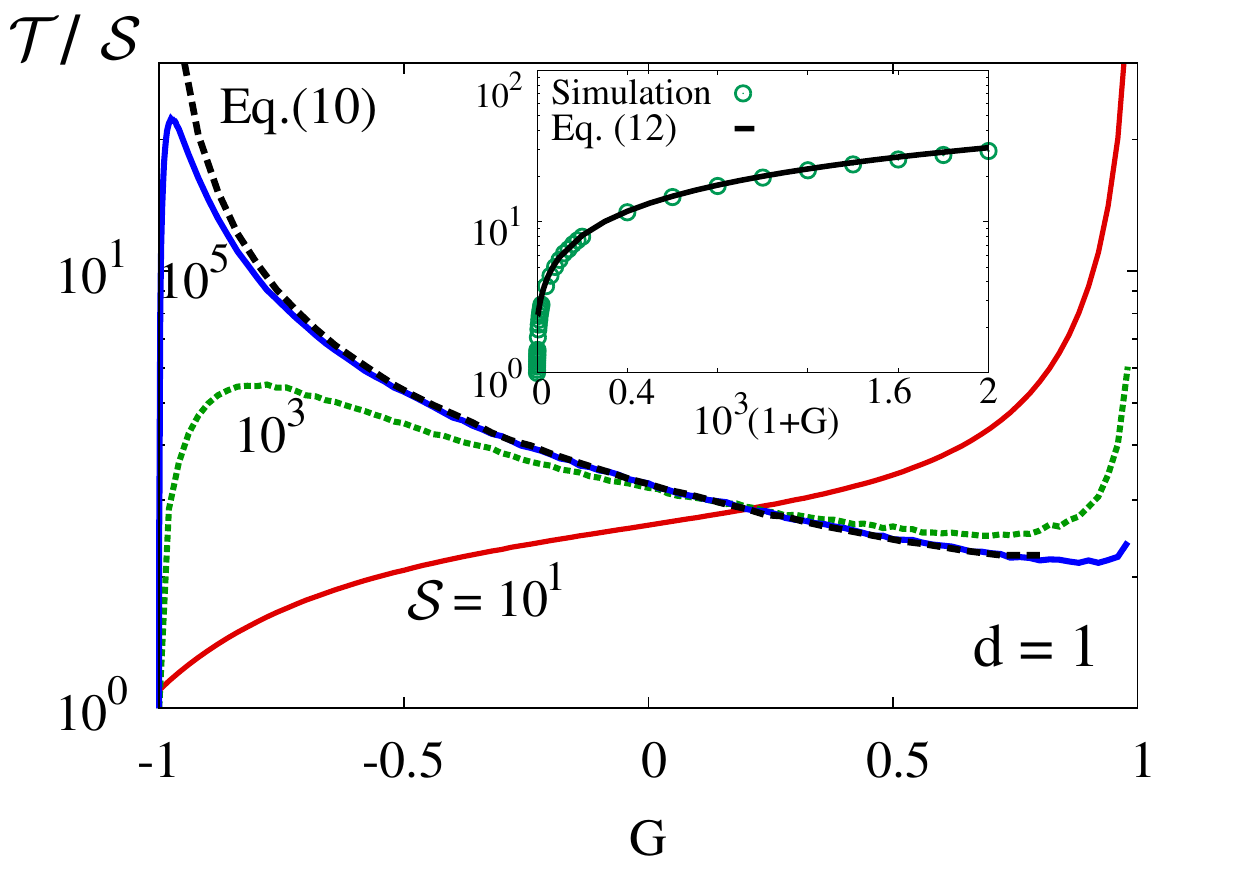}
\includegraphics[width=0.41\textwidth]{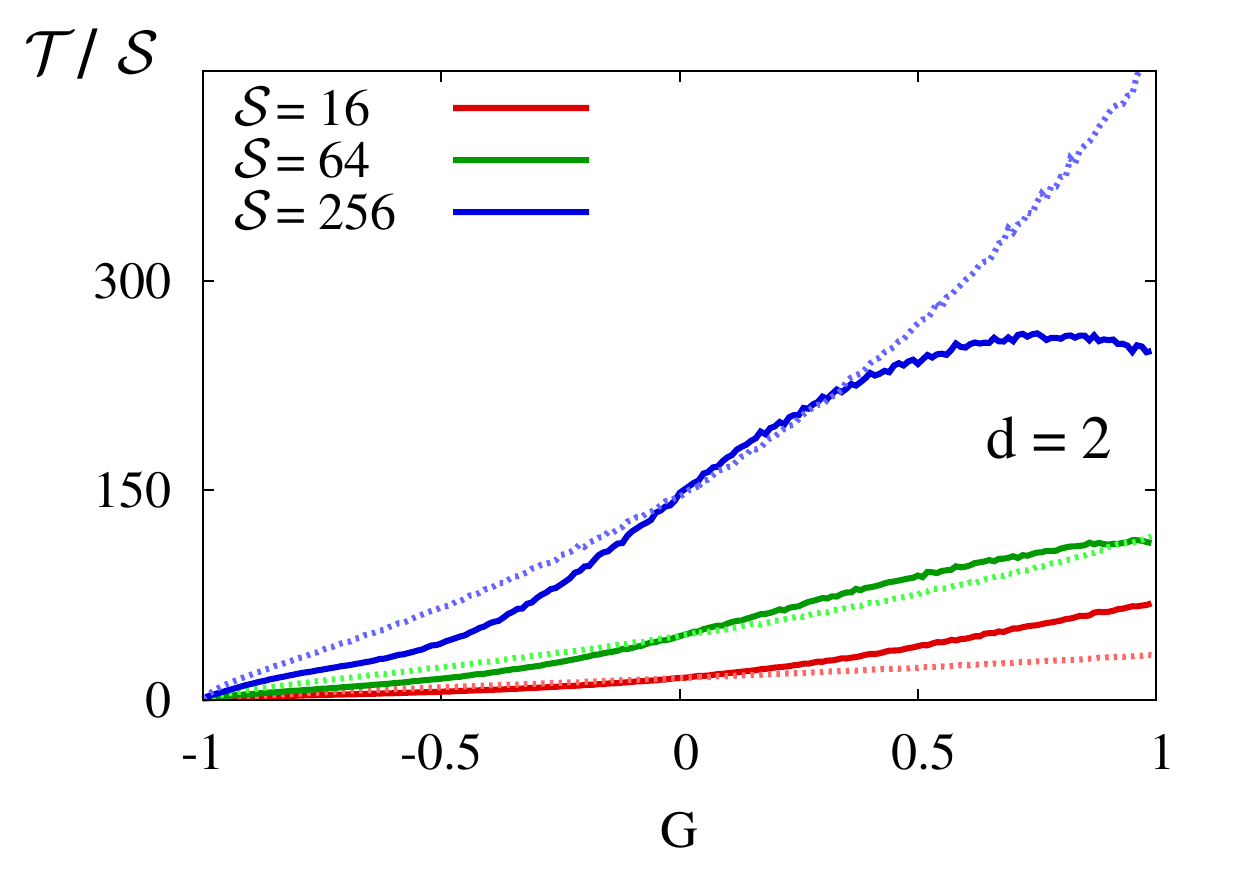}
\caption{ Dependence of the scaled forager lifetime $\mathcal{T}/\mathcal{S}$
  on greediness $G$ in $d=1$ and $d=2$.  The inset compares simulations with
  the analytic result \eqref{T-full} for $G$ close to $-1$ and
  $\mathcal{S}=10^6$.  Dotted curves in $d=2$ correspond to a non-backtracking
  walk (see summary text).}
\label{greed-prelim}
\end{figure}

We provide a heuristic argument that predicts both a non-monotonic dependence
of lifetime on greediness and a huge maximum for greediness $G\approx -1$
(Fig.~\ref{greed-prelim}).  Here, starvation proceeds in two stages: (i) The
forager first carves a critical desert of length $L_c$ by repeatedly reaching
either edge of the desert within $\mathcal{S}$ steps after food is consumed.
The critical length is defined by a forager of capacity $\mathcal{S}$
typically starving if it attempts to cross a desert of this length.  We
denote the time to create this critical-length desert as $\mathcal{T}_c$.
(ii) Once the desert length reaches $L_c$, the forager likely starves if it
attempts to cross the desert.  That is, the far side is unreachable and thus
irrelevant.  The time for this second stage is just the lifetime of a forager
in a semi-infinite desert, $\mathcal{T}_{\rm SI}$.

We now estimate the quantities $L_c$, $\mathcal{T}_c$, and
$\mathcal{T}_{\rm SI}$.  The time for a forager to reach food when it starts
a unit distance from food in a desert of length $k$ is given by
$t_1(k)=\frac{1-p}{p}\, k+3-\frac{2}{p}$ (see App.~\ref{app:escape}).
Therefore the time for the desert to grow to the critical length $L_c\gg 1$ is
\begin{equation}
\label{Tc}
  \mathcal{T}_c=\sum_{k=1}^{L_c} t_1(k)\simeq \frac{1-p}{p}\,\frac{L_c^2}{2}.
\end{equation}
We determine $L_c$ by equating the typical time to cross a desert of this
length, $t_\times \simeq \frac{2}{3} L_c^2+\frac{4L}{3p}$ (see App.~\ref{app:escape}), to
$\mathcal{S}$.  This gives two behaviors: $L_c\simeq \sqrt{3\mathcal{S}/2}$
for $p\gg 1/\sqrt{\mathcal{S}}$, and $L_c\simeq 3p\mathcal{S}/4$ for
$p\ll 1/\sqrt{\mathcal{S}}$.  Thus the time to reach the critical-length
desert is
\begin{equation}
\label{Ti}
\mathcal{T}_c\simeq
\begin{cases} \displaystyle{ 3(1\!-\!p)\mathcal{S}/4p} & \qquad p\gg
  1/\sqrt{\mathcal{S}}\,,\\[0.15cm]
 \displaystyle{9p\mathcal{S}^2/32} & \qquad p\ll 1/\sqrt{\mathcal{S}}\,.
\end{cases}
\end{equation}

\begin{figure}[ht]
\centerline{\includegraphics[width=0.325\textwidth]{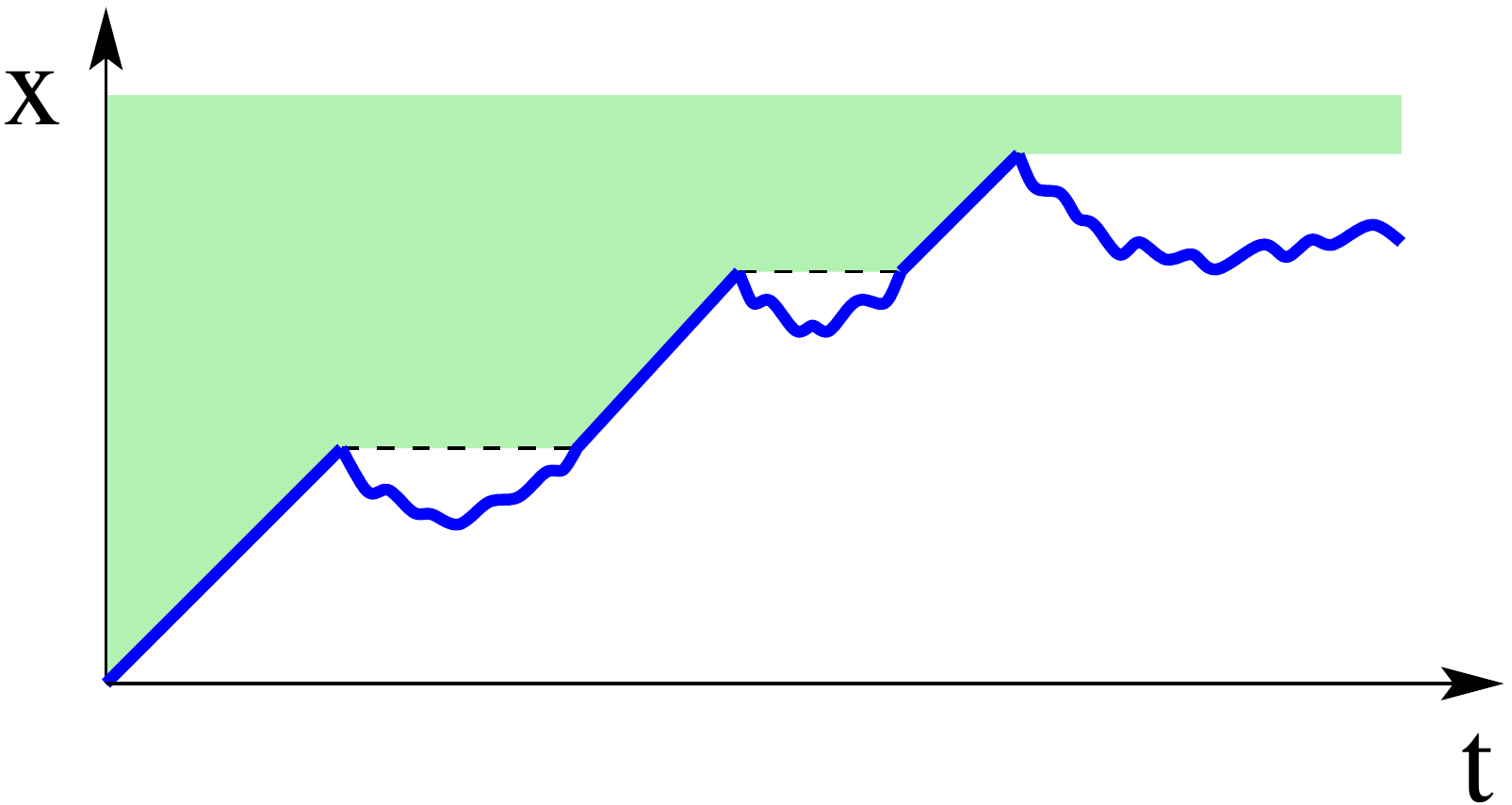}}
\caption{ Schematic illustration of the space-time trajectory of a greedy
  forager in the semi-infinite geometry.  The shaded region denotes food.}
\label{semi-inf-cartoon}
\end{figure}

For the semi-infinite geometry, a typical trajectory consists of segments
where the forager moves ballistically into the food-containing region,
interspersed by diffusive segments in the desert
(Fig~\ref{semi-inf-cartoon}).  As long as the diffusive segment lasts less
than $\mathcal{S}$ steps, the forager returns to the food/desert interface
and a new cycle of consumption and subsequent diffusion begins.  A ballistic
segment of $m$ consecutive steps towards food (followed by a step away)
occurs with probability $p^m(1-p)$.  The average time $t_b$ for this
ballistic segment is $t_b = \sum_{m\geq 1} m\, p^m\,(1\!-\!p)=p/(1\!-\!p)$.
The probability $\mathcal{R}$ for a diffusive segment to return to food
within $\mathcal{S}$ steps is the integral of the first-passage probability
for a forager that starts at $x=1$ to reach $x=0$ within time
$\mathcal{S}$~\cite{R01}:
\begin{align}
\mathcal{R}=\int_0^\mathcal{S} dt\,\,\frac{ e^{-1/4Dt}}{\sqrt{4\pi D t^3}} =
\mathrm{erfc}(1/\sqrt{4D\mathcal{S}})\,,\nonumber
\end{align}
where erfc$(\cdot)$ is the complementary error function.  The average number
of returns is
$\langle r \rangle= \sum_{r\geq 1} r \,\mathcal{R}^r(1\!-\!\mathcal{R})=
\mathcal{R}/(1\!-\!\mathcal{R})\simeq \sqrt{\pi\mathcal{S}/2}$
for $\mathcal{S}\to\infty$, where the asymptotics of the error function gives
the final result, and we take the diffusion coefficient $D=\frac{1}{2}$.  For
a forager that does return within $\mathcal{S}$ steps, the return time $t_r$
is thus
\begin{align}
  t_r  &=\frac{1}{\mathcal{R}} \int_0^\mathcal{S} dt\, t\, \frac{1}{\sqrt{4\pi Dt^3}}\,\,
e^{-1/4Dt} \simeq \sqrt{\frac{2\mathcal{S}}{\pi}} -1\,.\nonumber
\end{align}

The total trajectory therefore contains
$\langle r\rangle=\sqrt{\pi\mathcal{S}/2}$ elements, each of which are
comprised of a ballistic and a diffusive segment.  The time for each element
equals $t_b+t_r$.  There is also the final and fatal diffusive segment of
exactly $\mathcal{S}$ steps.  Consequently, the forager lifetime
$\mathcal{T}_{\rm SI}$ in the semi-infinite geometry is
\begin{align}
\label{T-semi}
\mathcal{T}_{\rm SI} &\simeq \langle r\rangle (t_b+t_r)+\mathcal{S}\simeq
\frac{2p-1}{1-p}\, \sqrt{\frac{\pi\mathcal{S}}{2}}+2\mathcal{S}\,.
\end{align}
From \eqref{Ti} and \eqref{T-semi}, we estimate the forager lifetime as
\begin{equation}
\label{T_qualit}
\mathcal{T} \simeq
\begin{cases}
\displaystyle{ \Big[\frac{3(1\!-\!p)}{4p} +2\Big] \mathcal{S}
+\frac{2p\!-\!1}{(1\!-\!p)}\,\sqrt{\frac{\pi\mathcal{S}}{2}}}\,,
& p\gg 1/\sqrt{\mathcal{S}}\,,\\[0.35cm]
\displaystyle{ \frac{9}{32}\,p\mathcal{S}^2+  2\mathcal{S
}} +\frac{2p\!-\!1}{(1\!-\!p)}\,
\sqrt{\frac{\pi\mathcal{S}}{2}}\,, & p\ll 1/\sqrt{\mathcal{S}}\,.
\end{cases}
\end{equation}

Two important consequences follow (Fig.~\ref{greed-prelim}):
\begin{itemize}
\item When $\mathcal{S}$ exceeds a critical value, it is easily seen that $\mathcal{T}$
  is decreasing with $p$, except for $p\to 0$ and $p\to 1$.  Since
  $\mathcal{T}$ diverges as $p\to 1$, the dependence of lifetime on
  greediness is non-monotonic!

\item For $p\simeq 1/\sqrt{\mathcal{S}}$, Eqs.~\eqref{T_qualit} give a common
  lifetime $\mathcal{T}\sim \mathcal{S}^{3/2}$---a huge maximum for large
  $\mathcal{S}$\,!  This maximum induces a second non-monotonicity in the
  negative greed (food averse) regime.
\end{itemize}

\section{One-Dimensional Solution}

We now outline the analytical solution for the forager lifetime that confirms
and quantifies the above heuristic picture.  The basic quantity is the
probability $V_k$ that the forager has eaten $k$ times at the instant of
starvation.  This quantity can be written as
\begin{equation}
\label{general_V}
V_k= \Big[1- \sum_{t=0}^\mathcal{S}\,  F_k(t)\Big]\,\,\prod_{j=1}^{k-1} \sum_{t=0}^\mathcal{S}\,  F_j(t)\,.
\end{equation}
Here $F_j(t)$ is the first-passage probability that a greedy forager that is
a unit distance from either edge of a desert of $k$ empty sites first reaches
either edge at time $t$.  The sum is thus the probability that this forager
escapes a desert of $j$ empty sites, and the product is the probability that
this forager successively escapes a desert of 1,2,3,\ldots, $k-1$ empty sites.
Finally the leading factor is the probability that the forager does not
escape a desert of $k$ empty sites.

We may now write the average forager lifetime as
\begin{align}
\label{general_T}
\mathcal{T}&=\sum_{k\geq 0}\, \big[ \sum_{j=1}^{k-1}\tau_j\big]\,V_k+\mathcal{S}\,.
\end{align}
Here
\begin{align*}
\tau_j=\frac{\sum_{0\leq t\leq\mathcal{S}}\,t \,F_j(t)}{\sum_{0\leq
  t\leq\mathcal{S}}\, F_j(t)}
\end{align*}
is the conditional average time for a greedy forager to successfully escape a
desert of $j$ empty sites when it starts one lattice spacing from either
edge.  The quantity $\sum_{j=1}^{k-1}\tau_j$ is the conditional time for the
forager to successively escape deserts of 1,2,3,\ldots, $k-1$ empty sites.
Consequently, the first term in \eqref{general_T} is that total time that the
forager takes to carve a desert of $k$ empty sites and the last factor,
$\mathcal{S}$, is the time for the last and fatal excursion in this desert.

To explicitly evaluate the forager lifetime in \eqref{general_T}, we need the
first-passage probability for a greedy forager, $F_k(t)$.  This first-passage
probability can be related to the unperturbed first-passage probability
$f_k(t)$ of a symmetric random walk by the convolution
\begin{equation}
\label{GLbasic}
F_k(t) = p\,\delta_{t,1} + (1-p)\sum_{t'\leq t-1} f_{k-2}(t')\,F_k(t-t'-1)\,.
\end{equation}
The first term accounts for a forager that reaches food in a single step.
The second term accounts for the forager hopping to the interior of the
interval.  In this case, the walker is at $x=2$ or $k-2$ and hops
symmetrically until it again reaches either $x=1$ or $k-1$.  Thus the
relevant first-passage probability is that for an unbiased random walk that
starts at $x=2$ or $k-2$ on $[1,k-1]$.  Once the walker first reaches either
$x=1$ or $k-1$, the process renews and the subsequent propagation involves
$F_k$.  Since one time unit is used in the first hop to the right, the walker
must reach the boundary in the remaining time $t-t'-1$ steps.
We solve Eq.~\eqref{GLbasic} by substituting in the generating functions
\begin{align}
{\widetilde f_k}(z)=\sum_{t\geq 1} f_k(t)\, z^t\,,\qquad
{\widetilde F_k}(z)=\sum_{t\geq 1} F_k(t)\, z^t\,.\nonumber
\end{align}
The generating functions reduce the convolution in Eq.~\eqref{GLbasic} to an
algebraic relation that is readily solved to give
\begin{equation}
\label{gzbasic}
{\widetilde F_k}(z) = \frac{pz}{1-(1-p)\,z \,{\widetilde f_{k-2}}(z)}\,.
\end{equation}

The next step is to substitute the well-known result for the Laplace
transform of the first-passage probability~\cite{R01}
\begin{align}
{\widetilde f_k}(s)&=
\text{sech}\sqrt{\frac{s}{D}}\,k \left[\sinh\Big(\sqrt{\frac{s}{D}}\Big)+
\sinh\Big(\sqrt{\frac{s}{D}}(k\!-\!1)\Big)\right]\,,\nonumber\\
&\underset{s\to 0}{\longrightarrow}
 1-\sqrt{\frac{s}{D}}\tanh\sqrt{\frac{sk^2}{4D}}k+\cdots\,.\nonumber
\end{align}
into Eq.~\eqref{gzbasic}.  We also convert the discrete generating function
to a continuous Laplace transform by replacing $z\to 1-s$.  This construction
is asymptotically exact in the limit $z\to 1$ or $s\to 0$, which corresponds
to the long-time limit in the time domain.  Following these steps, the
Laplace transform of the first-passage probability for the greedy forager for
$s\to 0$ and $k\to\infty$ is
\begin{equation}
\label{finite1}
{\widetilde F_k}(s)
=\bigg(1+\frac{1-p}{p}\sqrt{\frac{s}{D}}\tanh\sqrt{\frac{sk^2}{4D}}\bigg)^{-1}.
\end{equation}

Using the above first-passage probability for a greedy forager in a finite
desert, and also making use of standard Laplace transform manipulations, we
can determine both $\tau_k$ and $V_k$ in terms of ${\widetilde F_k}(s)$.
When these quantities are expressed in terms of
${\widetilde F_k}(s)$ in Eq.~\eqref{general_T}, we can finally determine the
forager lifetime $\mathcal{T}$.  These steps are somewhat tedious and all the
details are given in Ref.~\cite{BRB17}.  

There are two limiting cases where the forager lifetime has very different
asymptotic behaviors: $p\gg 1/\sqrt{\mathcal{S}}$ and
$p\ll 1/\sqrt{\mathcal{S}}$.  In the former case, we find
\begin{align}
\label{T-final}
\mathcal{T} &\simeq\mathcal{S}\,\frac{1\!-\!p}{p}\int_0^\infty \!\!\! d\theta\,
V_\theta\int_0^\theta \!\! \frac{du}{u}\sum_{j\geq 0}\frac{4}{v^2}\left\{
1\!-\!e^{-v^2}\!\left[1\!+\!v^2\right]\right\}+ \mathcal{S}\,.
\end{align}
Here $v=(2j\!+\!1)/u$, $\theta = n/(\pi \sqrt{D\mathcal{S}})$, with $n$ the
number of sites visited by the forager at starvation.  Additionally,
\begin{align}
\label{V-final}
V_\theta &\simeq \frac{4(1\!-\!p)}{p\theta}\sum_{j\geq 0} e^{-w^2-Q}\,,\quad
Q= \frac{2(1-p)}{p}\sum_{j\geq 0} E_1(w^2)\nonumber\,,
\end{align}
where $w=(2j\!+\!1)/\theta$, and
$E_1(x)=\int_1^\infty \frac{dt}{t}\, e^{-xt}$ is the exponential integral.
Because the function $V_\theta$ depends on $p$, the greedy forager lifetime
$\mathcal{T}$ does \emph{not} merely equal $\mathcal{T}$ for the non-greedy
forager times $\frac{1-p}{p}$.  Our result~\eqref{T-final} agrees with
numerical simulations for large $\mathcal{S}$ (Fig.~\ref{greed-prelim}).

Deep in the negative greed regime $p\ll 1/\sqrt{\mathcal{S}}$,
Eq.~\eqref{finite1} simplifies to 
\begin{equation}
{\widetilde F}_k(s)\simeq \left(1+\frac{k}{2pD}s\right)^{-1}\,.
\end{equation}
Following the same steps as given above now leads to
\begin{equation}
\label{T-full}
\mathcal{T}=  \sum_{k\geq 1} \frac{k^2}{2p \mathcal{S}}\, e^{-{2pD\mathcal{S}}/{k}}
\exp\left[-\int_1^k e^{-{2pD\mathcal{S}}/{x}}\,dx\right]+\mathcal{S}\,,
\end{equation}
whose numerical evaluation matches the simulation results in the regime
$p\ll 1/\sqrt{\mathcal{S}}$ (Fig.~\ref{greed-prelim} inset).

\section{Two Dimensions} 

\begin{figure}[H]
  \centerline{\includegraphics[width=0.4\textwidth]{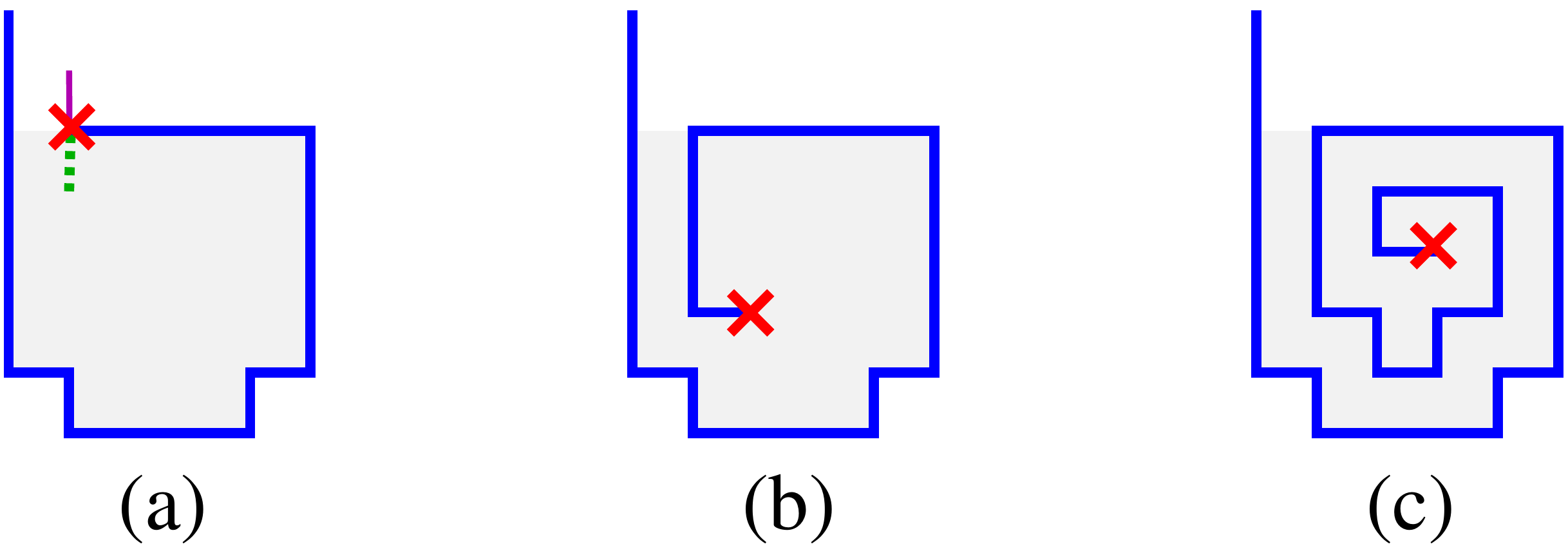}}
  \caption{ A random-walk trajectory that leads to trapping of a perfectly
    greedy forager. (a) Forager ($\times$) at the decision point.  (b)
    Forager hops to the interior region (shaded).  (c) Food in the interior
    is completely consumed, so that the forager ($\times$) may be trapped
    inside the newly created desert.}
\label{moat}
\end{figure}

Surprisingly, simulations show that the forager lifetime again varies
non-monotonically with (positive) greed, but in the opposite sense compared
to one dimension (Fig.~\ref{greed-prelim}).  A perfectly greedy forager has a
\emph{smaller} lifetime than one that is not quite as avaricious.  We can
explain this feature in a simple way: Because a random walk is recurrent in
two dimensions, it will certain form closed loops along its
trajectory~\cite{F68,W94}.  Suppose that a perfectly greedy forager is about
to form such a closed loop (Fig.~\ref{moat}(a)).  At this point, the forager
has only two possible choices for the next step.  One of them leads outside
the incipient closed loop and the other leads inside.  If the latter choice
is made, a ``moat'' is created by the previous trajectory.

Once inside the moat, a perfectly greedy forager always consumes food in its
nearest neighborhood.  Ultimately, this interior food is mostly or completely
depleted (the latter is shown in Fig.~\ref{moat}(c)).  While the former case
is more likely, the remaining food will be scarce and isolated.  Thus the
forager creates and then becomes trapped inside a (perhaps slightly
imperfect) desert.

Conversely, if the greediness $G<1$, a forager that encounters the moat from
the interior can cross it with a non-zero probability and thereby reach food
on the outside.  This mechanism provides a route for the forager to escape
the desert and survive longer than if it remained strictly inside.  This
argument indicates that the forager lifetime should be a decreasing function
of $G$ as $G\to 1$, as confirmed by simulations (Fig.~\ref{greed-prelim}).
Also in stark contrast to one dimension, there is no peak in the forager
lifetime for negative greed, at least for the values of $\mathcal{S}$ that we
were able to simulate.

\section{Summary}

Greed plays a paradoxical role in the lifetime of a greedy random-walking
forager, which moves preferentially towards local food for positive
greediness, and away from food for negative greediness.  The lifetime depends
\emph{non-monotonically} on greediness when the forager capacity is
sufficiently large.  Moreover, the sense of the non-monotonicity is opposite
in one and two dimensions.  In $d=1$, the forager lifetime exhibits a huge
peak of the order of $S^{3/2}$ for $G\approx -1$, scales as
$\mathcal{S}^{1/2}/(1-G)$ for $G\to 1$, while $\mathcal{T}\simeq \mathcal{S}$
throughout the rest of the range of $G$.  Determining these intriguing
properties rests on solving a challenging non-Markovian first-passage problem
in which the forager motion is locally biased when food is in the forager's
nearest neighborhood.

A variety of questions remain open.  Can one make analytical progress in two
dimensions?  What is the behavior of the lifetime in greater than two
dimensions?  Simulations are not useful here because the lifetime is
extremely long for non-negligible greed and memory/computation time
constraints become prohibitive.  On a biological note, greed can be viewed as
endowing a forager with a minimal information processing capability.  A
related mechanism is for the forager to perform a non-backtracking random
walk (previous step is not retraced).  The forager lifetime increases
monotonically with the probability of not backtracking
(Fig.~\ref{greed-prelim}; here $1-G$ is a proxy for the backtracking
probability) and perfect non-backtracking is superior to perfect greed.  It
would be useful to understand how to most effectively increase the forager
lifetime with minimal information-processing enhancements to random-walk
motion.

We acknowledge support from the European Research Council starting grant
FPTOpt-277998 (OB), from grants DMR-1608211 and DMR-1623243 from the National
Science Foundation (UB and SR), by the John Templeton Foundation (SR), and
from grant 2012145 from the U.S.-Israel Binational Science Foundation (UB).


\appendix
\section{Escape From An Interval}
\label{app:escape}

\begin{figure}[ht]
\centerline{\includegraphics[width=0.45\textwidth]{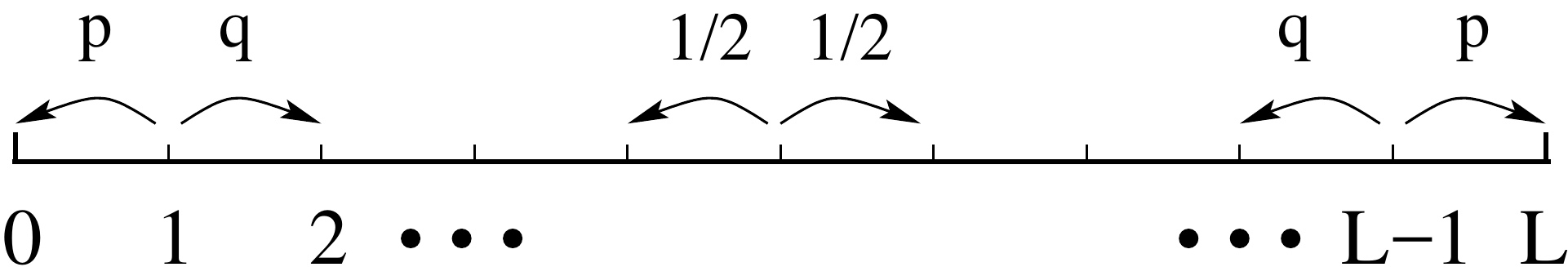}}
\caption{ Hopping probabilities for a greedy forager inside a desert of
  length $L$.}
\label{interval}
\end{figure}

We determine the first-passage properties of a random walk in a finite
interval of length $L$ whose hopping rules are the same as that of a greedy
forager.  That is, a walk in the interior hops equiprobably to the left and
right, while a walk at either $x=1$ or $x=L-1$ hops to the edge of the
interval with probability $p$ and into the interior with probability $1-p$
(Fig.~\ref{interval}).  For these hopping rules, we calculate the exit
probabilities to each side of the interval, the unconditional time to exit
either side of the interval, and the conditional exit time to exit by each
edge of the interval.  We will use the result for the unconditional exit time
to derive Eq.~\eqref{Tc}, from which we will heuristically argue that the
lifetime of a forager with a sufficiently large capacity varies
non-monotonically with greediness.

Let $E_n$ be the probability that the forager, which starts at site $n$,
exits the interval via the left edge.  The exit probabilities satisfy the
backward equations
\begin{align}
\begin{split}
&E_1= p +q E_2\,, \\
&E_n = \tfrac{1}{2} E_{n-1}+\tfrac{1}{2} E_{n+1} \qquad 2\leq n\leq L-2\,,\\
&E_{L-1} = qE_{L-2}\,.
\end{split}
\end{align}
No boundary conditions are needed, as the distinct equations for $n=1$ and
$n=L-1$ fully determine the exit probabilities.  As we shall see, $E_n=0$ not
at $n=L$, but at different value of $n$, and similarly for the point where
$E_n=1$.

Since the deviation to random-walk motion occurs only at the boundaries, we
attempt a solution that has the random-walk form in the interior of the
interval: $E_n=A+Bn$.  This ansatz automatically solves the interior
equations ($2\leq n\leq L-2$), while the boundary equations for $n=1$ and
$n=L-1$ give
\begin{align*}
E_1=p+qE_2 &~\longrightarrow~ A+B=p+q(A+2B)\,,\\
E_{L-1}=qE_{L-2} &~\longrightarrow~ A+B(L-1)=q\big(A+B(L-2)\big)\,,
\end{align*}
from which $A$ and $B$ are
\begin{equation*}
A=\frac{p(L-2)+1}{pL+2(1-2p)}\,,\qquad B=-\frac{p}{pL+2(1-2p)}\,.
\end{equation*}
Thus the probability that a greedy random walk that starts at $x=n$ exits via
the left edge of the interval is
\begin{equation}
\label{En}
  E_n=A+Bn= \frac{L-n+\frac{1}{p}(1-2p)}{L+\frac{2}{p}(1-2p)}\,,
\end{equation}
while the exit probability via the right edge is $1-E_n$.  As might be
expected for a perturbation that applies only at the boundary, the overall
effect of greed on the exit probability is small: the exit probability
changes from $E_n=1-\frac{n}{L}$ for $p=\frac{1}{2}$ to
$E_n=1-\frac{n-1}{L-2}$ for $p=1$.  That is, the effective interval length
changes from $L$ to $L-2$ as $p$ increases from $\frac{1}{2}$ to 1.


Similarly, let $t_n$ be the average time for a greedy random walker to reach
either edge of the interval when the walk starts at site $n$.  These exit
times satisfy the backward equations
\begin{align}
\begin{split}
\label{tn}
&t_1= p +q(t_2+1)\,, \\
&t_n = \tfrac{1}{2} t_{n-1}+\tfrac{1}{2} t_{n+1} +1\qquad 2\leq n\leq L-2\,,\\
&t_{L-1} = p+q(t_{L-2}+1)\,.
\end{split}
\end{align}
Again, no boundary conditions are needed, as the equations for $n=1$ and
$n=L-1$ are sufficient to solve \eqref{tn}.  We attempt a solution for these
second-order equations that has the same form as in the case of no greed:
$t_n= a+bn+cn^2$.  Substituting this ansatz into \eqref{tn} immediately gives
$c=-1$, while the equations for $t_1$ and $t_{L-1}$ lead to the conditions
\begin{align}
&-1+a+b = q(-4+a+2b)+1\,,\nonumber \\
&-(L-1)^2+b(L-1)+a \nonumber \\
&\qquad\qquad~~~ = q\big[-(L-2)^2+b(L-2)+a\big]+1\,.\nonumber
\end{align}

Solving these equations, the average exit time to either edge of the interval
when starting from site $n$ is
\begin{equation}
\label{tn:app}
t_n=n(L-n)-\frac{2p-1}{p}(L-2)\,.
\end{equation}
This gives a parabolic dependence of $t_n$ on $n$ that is shifted slightly
downward compared to the case of no greed, as $p$ ranges from $\frac{1}{2}$
to 1.  Notice again that $t_n=0$ not at $n=0$ and $n=L$, but rather at points
between $n=0$ and 1 and between $n=L-1$ and $L$ for $p>\frac{1}{2}$.  This
overall shift leads to a tiny change in each $t_n$, \emph{except} when the
forager starts one site away from the boundary.   

Finally, we determine the \emph{conditional} exit
times, $t^\pm_n$, defined as the time to reach left edge of the interval when
starting from site $n$ (for $t^-$) and to the right edge (for $t^+$),
conditioned on the walker exiting only by the specified edge.  We focus on
$t^-_n$, because once $t^-_n$ is determined, we can obtain $t^+_n$ via
$t^+_n=t^-_{L-n}$.  The conditional exit times $t^-_n$ satisfy
\begin{align}
\begin{split}
\label{un}
&u_1= qu_2+E_1\,, \\
&u_n = \tfrac{1}{2} u_{n-1}+\tfrac{1}{2} u_{n+1} +E_n\qquad 2\leq n\leq L-2\,,\\
&u_{L-1} = qu_{L-2}+E_{L-1}\,,
\end{split}
\end{align}
where $u_n\equiv E_nt^-_n$, with $E_n$, the exit probability to the left
edge, given by Eq.~\eqref{En}.  Because Eqs.~\eqref{un} are second-order with
an inhomogeneous term that is linear in $n$, the general solution is a cubic
polynomial: $u_n=a+bn+cn^2+dn^3$.  Substituting this form into Eq.~\eqref{un}
for $2\leq n\leq L-2$, we obtain the conditions $c=-A$ and $d=-B/3$, where
$A$ and $B$ are the coefficient of $E_n$ in Eq.~\eqref{En}.  The remaining
two coefficients are determined by solving the equations for $u_1$ and
$u_{L-1}$ and the final results for the coefficients $a,b,c,d$ in $u_n$ are:
\begin{align}
\label{abcd}
a&=\frac{2 (L\!-\!2) (1\!-\!2p) \big[p^2(L\!-\!4)(L+\frac{3}{p}(1\!-\!p))+3\big]}{3 p^3 
\big[L+\frac{2}{p}(1-2p)\big]^2}\,,\nonumber\\
b&=\frac{2p^2 \big[L(L^2\!-\!6L\!+\!6)\!+\!8\big]+6pL(L\!-\!3) \!+\!6L\!-\!8 p}{3 p^2 \big[L+\frac{2}{p}(1-2p)\big]^2}\,,\nonumber\\
c&=-\frac{L+\frac{1}{p}(1-2p)}{L+\frac{2}{p}(1-2p)}\,, \nonumber\\
d&=-\frac{1}{3}\,\frac{1}{L+\frac{2}{p}(1-2p)}\,.
\end{align}
Finally, the conditional exit time to the left edge is $t_n^-=u_n/E_n$, with
$u_n=a+bn+cn^2+dn^3$, and $E_n$ already determined in Eq.~\eqref{En}.  We are
particularly interested in $t^-_{L-1}$, the conditional time for a walk that
starts at $x=L-1$ to reach $x=0$.  From Eqs.~\eqref{En} and \eqref{abcd}, the
limiting behavior of this crossing time for large $L$ is
\begin{equation}
\label{app:tx}
t^-_{L-1}\equiv t_\times\simeq \frac{2}{3}\,L^2 + \frac{4}{3}\,\frac{L}{p}\,,
\end{equation}

\end{document}